\documentclass[showpacs,preprintnumbers,onecolumn,superscriptaddress,aps,nofootinbib]{revtex4}
\usepackage{amssymb}
\usepackage[centertags]{amsmath}
\usepackage{txfonts}
\usepackage{epsfig}
\usepackage{bm}
\usepackage{color}
\usepackage{graphicx,graphics}
\usepackage{multirow}
\usepackage{float}
\usepackage{ulem}
\usepackage{enumerate}
\usepackage[pdfstartview=FitH]{hyperref}
\allowdisplaybreaks[2]

\begin{document}

\title{Hidden-charm Pentaquark States in Heavy Ion Collisions at the Large Hadron Collider}

 \author{Rui-Qin Wang}
 \affiliation{College of Physics and Engineering, Qufu Normal University, Shandong 273165, China}
 \affiliation{Department of Physics and Astronomy and Shanghai Key Laboratory for Particle Physics and Cosmology, Shanghai Jiao Tong University, Shanghai 200240, China}
 
\author{Jun Song}
\affiliation{Department of Physics, Jining University, Shandong 273155, China}

\author{Kai-Jia Sun}
\affiliation{Department of Physics and Astronomy and Shanghai Key Laboratory for Particle Physics and Cosmology, Shanghai Jiao Tong University, Shanghai 200240, China}

\author{Lie-Wen Chen}
\affiliation{Department of Physics and Astronomy and Shanghai Key Laboratory for Particle Physics and Cosmology, Shanghai Jiao Tong University, Shanghai 200240, China}

\author{Gang Li}
\affiliation{College of Physics and Engineering, Qufu Normal University, Shandong 273165, China}

\author{Feng-Lan Shao}
\email {shaofl@mail.sdu.edu.cn}
\affiliation{College of Physics and Engineering, Qufu Normal University, Shandong 273165, China}

\begin{abstract}

In the framework of the quark combination, we derive the yield formulas and study the yield ratios of the hidden-charm pentaquark states 
in ultra-relativistic heavy ion collisions.
We propose some interesting yield ratios which clearly exhibit the production relationships between different hidden-charm pentaquark states.
We show how to employ a specific quark combination model to evaluate the yields of exotic $P_c^+(4380)$, $P_c^+(4450)$ and their partners
on the basis of reproducing the yields of normal identified hadrons, 
and execute the calculations in central Pb+Pb collisions at $\sqrt{s_{NN}}= 2.76$ TeV as an example.
 
\end{abstract}

\pacs{25.75.Dw, 25.75.Nq, 25.75.-q}
\maketitle

\section{introduction}

The LHCb Collaboration at the Large Hadron Collider (LHC) has recently announced two exotic resonances $P_c^+(4380)$ and $P_c^+(4450)$, 
consistent with pentaquark states, via the $J/\psi p$ invariant mass spectrum in $\Lambda_b^0$ decays in $p+p$ reactions \cite{PcLHCb2015PRL}.
Thereupon, the study for the internal configurations of such resonances becomes a subject of an intensive discussion in literatures.
The so-far suggested interpretations for $P_c^+(4380)$ and $P_c^+(4450)$ include molecular states bound of a charmed baryon 
and an anticharmed meson \cite{molecular0,molecular1},
molecular states with $J/\psi$ and an excited nucleon \cite{molecular2},
composites containing colored baryonlike and mesonlike constituents \cite{colormolecular}, 
pentaquark states with one heavy-light diquark, one light-light diquark and a charm antiquark \cite{diquark1,diquark2,diquark3},
pentaquark states with color-antitriplet diquark $cu$ and color-triplet $\bar c ud$ \cite{cu-cbud1,cu-cbud2}, and so on.
There are also some works about needing further confirmation of whether such resonance is just a kinematical effect or a real exotic resonance
by analyzing other processes, such as $\Lambda_b^0\rightarrow K^- \chi_{c1} p$ \cite{nomolecular3,WWangPRD2015,confirm3,confirm4}.
Further exploring for the intrinsic dynamical nature of such exotic resonances and searching for their partners are necessary for understanding 
some longstanding questions in hadronic physics, which requires more theoretical and experimental efforts.

Heavy ion collisions with ultra-relativistic collision energies, especially those at the LHC, 
provide preferable conditions for the creation of particles in heavy flavor sectors. 
The studies of heavy flavor exotic resonances in heavy ion collisions not only are complementary to those in the LHCb experiments, 
but also offer unique insights into some of the fundamental questions in hadronic physics \cite{ExHIC2011PRL,ExHIC2011PRC}.
As is well known, the Quark Combination Mechanism (QCM) is an effective phenomenological method to deal with the hadronization of the partonic
system produced early in high energy heavy ion collisions. It has shown successes in reproducing multiplicities, yield ratios, momentum distributions,
elliptic flows, etc., of normal identified light, strange and heavy flavor hadrons \cite{ALCOR1995PLB,Zimanyi2000PLB,Fries2003,Greco2003,DAAPLB2004,TYao2008PRC,RlamcD0com}, 
and has also many applications in describing the production properties of exotic resonances \cite{ExHIC2011PRL,ExHIC2011PRC,LWC2004PLB,FLShao2005PRC,LWChen2007PRC,Lee2008EPJC}.
Due to the produced bulk deconfined fireball and the hadronization through the quark combination in high energy heavy ion collisions, 
various kinds of exotic hadrons can be formed.
The purpose of this paper is to investigate the yield ratios and estimate the yields of various hidden-charm pentaquark states 
with different valence quark ingredients and different intrinsic quantum numbers in the framework of the QCM 
in ultra-relativistic heavy ion collisions at the LHC. 
This can provide useful references for the exotic hadron search in future experiments and is helpful for understanding
the production mechanism of the exotic resonances in heavy ion collisions.

The rest of the paper is organized as follows. 
In Sec.~II, we derive the yield formulas and study the yield ratios of hidden-charm pentaquark states from the basic ideas of the QCM.
During the derivation, a few assumptions, approximations and/or simplifications are used and they are all clearly presented.
In Sec.~III, we show how to employ a specific quark combination model to estimate the yields of different hidden-charm pentaquark states,
and give the estimated results in central Pb+Pb collisions at $\sqrt{s_{NN}}= 2.76$ TeV as an example. 
Sec.~IV summaries our work.

\section{Yield ratios of different hidden-charm pentaquark states}

In this section, we derive the yield formalism of the hidden-charm pentaquark states in the quark combination models based on the basic ideas.
We first begin with a quark-antiquark system as general as possible. Then we simplify the results by adopting a few explicit assumptions and/or
approximations. Finally we present some interesting results for the yield ratios of different hidden-charm pentaquark states.

\subsection{The yield formalism of the hidden-charm pentaquark states}

We start from a color-neutral quark-antiquark system with $N_{q_i}$ quarks of flavor $q_i$ ($q_i=u,~d,~s,~c$) and $N_{\bar q_i}$ antiquarks of 
flavor $\bar q_i$ ($\bar q_i=\bar u,~\bar d,~\bar s,~\bar c$). 
All these quarks and antiquarks can hadronize via the quark combination into not only normal mesons, baryons and antibaryons, 
but also exotic hadrons such as tetraquark states, pentaquark states and so on.
The momentum distribution $f_{P_j}(p;N_{q_i},N_{\bar{q}_i})$ for the directly produced pentaquark state $P_j$ with the known quark
contents $(q_{01}q_{02}q_{03}q_{04}\bar q_{05})$ after quark combination hadronization is given by
{\setlength\arraycolsep{0.2pt}
\begin{eqnarray}
f_{P_j}(p;N_{q_i},N_{\bar{q}_i})= \sum\limits_{q_{1}q_{2}q_{3}q_{4}\bar q_{5}}  \int dp_1 dp_2 dp_3 dp_4 dp_5  
                                       f_{q_{1}q_{2}q_{3}q_{4}\bar q_{5}}(p_1,p_2,p_3,p_4,p_5;N_{q_i},N_{\bar{q}_i})  
                       \mathcal {R}_{P_j,q_{1}q_{2}q_{3}q_{4}\bar q_{5}}(p,p_1,p_2,p_3,p_4,p_5;N_{q_i},N_{\bar{q}_i}), 
                       \nonumber   \\   \label{eq:fPjgeneral}
\end{eqnarray} }%
where $f_{q_{1}q_{2}q_{3}q_{4}\bar q_{5}}$ is the five-particle joint momentum distribution for $(q_{1}q_{2}q_{3}q_{4}\bar q_{5})$.
The kernel function $\mathcal {R}_{P_j,q_{1}q_{2}q_{3}q_{4}\bar q_{5}}$ stands for the probability density for $q_1$, $q_2$, $q_3$, $q_4$, and $\bar{q}_5$ 
with momenta $p_1$, $p_2$, $p_3$, $p_4$, and $p_5$ to combine into a pentaquark state $P_j$ of momentum $p$. 
Integrating over $p$ from Eq.~(\ref{eq:fPjgeneral}), we can obtain the average number of the directly produced $P_j$ as
{\setlength\arraycolsep{0.2pt}
\begin{eqnarray}
\overline N_{P_j}(N_{q_i},N_{\bar{q}_i})= \sum\limits_{q_{1}q_{2}q_{3}q_{4}\bar q_{5}}  \int dp dp_1 dp_2 dp_3 dp_4 dp_5 
                                          f_{q_{1}q_{2}q_{3}q_{4}\bar q_{5}}(p_1,p_2,p_3,p_4,p_5;N_{q_i},N_{\bar{q}_i})  
                       \mathcal {R}_{P_j,q_{1}q_{2}q_{3}q_{4}\bar q_{5}}(p,p_1,p_2,p_3,p_4,p_5;N_{q_i},N_{\bar{q}_i}).
                       \nonumber \\  \label{eq:NPjgeneral}
\end{eqnarray} }%
Eqs.~(\ref{eq:fPjgeneral}) and (\ref{eq:NPjgeneral}) are the most general starting point of describing the production of pentaquark states with
any flavors of quark ingredients in high energy reactions based on the basic ideas of the QCM. 
Different models are some special examples of the general case we consider in these equations. In specific models, different methods and/or 
assumptions are introduced to construct the precise form of the kernel function.
For example, the kernel function evolves into the Wigner function in the coalescence model \cite{Greco2003} and 
the recombination function in the quark recombination model \cite{Hwa2003PRC}, respectively.

For a special kind of hidden-charm pentaquark states $P^{c\bar c}_j$ with the known quark contents $(l_{01}l_{02}l_{03}c\bar c)$ 
distinguished by the superscript $c\bar c$, we from Eq.~(\ref{eq:NPjgeneral}) easily have
{\setlength\arraycolsep{0.2pt}
\begin{eqnarray}
\overline N_{P^{c\bar c}_j}(N_{q_i},N_{\bar{q}_i})= \sum\limits_{l_{1}l_{2}l_{3}}  \int dp dp_1 dp_2 dp_3 dp_4 dp_5 
                                          f_{l_{1}l_{2}l_{3}c\bar c}(p_1,p_2,p_3,p_4,p_5;N_{q_i},N_{\bar{q}_i})  
                       \mathcal {R}_{P_j^{c\bar c},l_{1}l_{2}l_{3}c\bar c}(p,p_1,p_2,p_3,p_4,p_5;N_{q_i},N_{\bar{q}_i}),
                         \label{eq:NPccbgeneral}
\end{eqnarray} }%
where $l_i=u,~d,~s$.
The joint momentum distribution $f_{l_{1}l_{2}l_{3}c\bar c}$ is the number density that satisfies
{\setlength\arraycolsep{0.2pt}
\begin{eqnarray}
 \int dp_1 dp_2 dp_3 dp_4 dp_5  f_{l_{1}l_{2}l_{3}c\bar c}(p_1,p_2,p_3,p_4,p_5;N_{q_i},N_{\bar{q}_i})  
 =N_{l_{1}l_{2}l_{3}c\bar c}  = N_{l_{1}l_{2}l_{3}} N_c N_{\bar c},                     
\end{eqnarray} }%
where $N_{l_{1}l_{2}l_{3}c\bar c}$ and 
\begin{equation}
N_{l_1l_2l_3} = \left\{ \begin{array}{ll}
N_{l_1}N_{l_2}N_{l_3} & \textrm{for    } l_1 \neq l_2 \neq l_3\\
N_{l_1}(N_{l_1}-1)N_{l_3} & \textrm{for    } l_1=l_2 \neq l_3,\\
N_{l_1}(N_{l_1}-1)(N_{l_1}-2) & \textrm{for    } l_1=l_2=l_3
\end{array} \right.
\end{equation}
are the numbers of all the possible $(l_{1}l_{2}l_{3}c\bar c)$'s and $(l_{1}l_{2}l_{3})$'s, respectively, in the considered bulk quark-antiquark system.
We rewrite
{\setlength\arraycolsep{0.2pt}
\begin{eqnarray}
 f_{l_{1}l_{2}l_{3}c\bar c}(p_1,p_2,p_3,p_4,p_5;N_{q_i},N_{\bar{q}_i})  
 =N_{l_{1}l_{2}l_{3}c\bar c} f^{(n)}_{l_{1}l_{2}l_{3}c\bar c}(p_1,p_2,p_3,p_4,p_5;N_{q_i},N_{\bar{q}_i}),                     
\end{eqnarray} }%
so that the joint momentum distribution is normalized to one where is denoted by the superscript $(n)$, i.e.,
{\setlength\arraycolsep{0.2pt}
\begin{eqnarray}
 \int dp_1 dp_2 dp_3 dp_4 dp_5  f^{(n)}_{l_{1}l_{2}l_{3}c\bar c}(p_1,p_2,p_3,p_4,p_5;N_{q_i},N_{\bar{q}_i})  =1.                    
\end{eqnarray} }%
We adopt an assumption of $u,~d,~s$-flavor independence of the normalized joint momentum distribution of quarks and/or antiquarks, i.e.,
{\setlength\arraycolsep{1pt}
\begin{eqnarray}
f^{(n)}_{l_{1}l_{2}l_{3}c\bar c}(p_1,p_2,p_3,p_4,p_5;N_{q_i},N_{\bar{q}_i})= f^{(n)}_{lllc\bar c}(p_1,p_2,p_3,p_4,p_5;N_l,N_{\bar l},N_q,N_{\bar q}),
\end{eqnarray}}%
to simplify the formalism. Here, $l$ stands for $u$, $d$ or $s$; $N_l$ and $N_{\bar l}$ stand for the total number of $u,~d$ and $s$ quarks and 
that of $\bar u,~\bar d$ and $\bar s$ antiquarks, respectively; 
$N_q$ and $N_{\bar q}$ stand for the total number of $u,~d,~s,~c$ quarks and 
that of $\bar u,~\bar d,~\bar s,~\bar c$ antiquarks in the considered quark-antiquark system.
With the normalized $u,~d,~s$-flavor independent joint momentum distribution, we have
{\setlength\arraycolsep{0.2pt}
\begin{eqnarray}
\overline N_{P^{c\bar c}_j}(N_{q_i},N_{\bar{q}_i})= \sum\limits_{l_{1}l_{2}l_{3}}  N_{l_{1}l_{2}l_{3}c\bar c}  \int dp dp_1 dp_2 dp_3 dp_4 dp_5 
                                          f^{(n)}_{lllc\bar c}(p_1,p_2,p_3,p_4,p_5;N_{q_i},N_{\bar{q}_i})  
                 \nonumber   \\
           \times      \mathcal {R}_{P_j^{c\bar c},l_{1}l_{2}l_{3}c\bar c}(p,p_1,p_2,p_3,p_4,p_5;N_{q_i},N_{\bar{q}_i}).
                         \label{eq:NPccb-assum1}
\end{eqnarray} }%

The kernel function $\mathcal {R}_{P_j^{c\bar c},l_{1}l_{2}l_{3}c\bar c}$ carries a lot of kinematical and dynamical information on the quark combination hadronization,
but its precise form is very ambiguous at present due to its complicated non-perturbative property.
Despite all this, we have known the kernel function should be endowed with the following four constraints at least. 
The first is satisfying the momentum conservation, so it should contain the item $\delta(\sum^5_{i=1} p_i-p)$. 
The second is the quark flavor conservation during the hadronization process, which is the requirement of the flavor conservation in the strong interaction.
The third is the requirement of the hadronization unitarity, i.e., the production of normal and exotic hadrons should exhaust all quarks and antiquarks in the system after hadronization.
The fourth is the dynamics of hadron-hadron production competition, i.e., when a quark hadronizes, whether
it forms a meson by combining an antiquark or forms a baryon by combining other two quarks, or goes into an exotic state as other constraints are all satisfied in these cases.
Similar detailed discussions of kernel functions on normal hadrons can be found in our previous work \cite{RQWang2015PRC}. 
Based on the above four points, we assume a simple but explicit case where the kernel function can be factorized as follows
{\setlength\arraycolsep{0.2pt}
\begin{eqnarray}
\mathcal {R}_{P_j^{c\bar c},l_{1}l_{2}l_{3}c\bar c}(p,p_1,p_2,p_3,p_4,p_5;N_{q_i},N_{\bar{q}_i})
=C_{P^{c\bar c}_j}
 \mathcal {R}_{l_{01}l_{02}l_{03}} 
 \mathcal {R}_{P^{c\bar c},l_1l_2l_3c\bar c}(p_1,p_2,p_3,p_4,p_5;N_{q_i},N_{\bar{q}_i})
 \delta(\displaystyle{\sum^5_{i=1}} p_i-p),  \label{eq:Rfact}
\end{eqnarray} }%
where the $\delta$ function guarantees the momentum conservation.
$\mathcal {R}_{P^{c\bar c},l_1l_2l_3c\bar c}$ denotes the probability of $l_1$, $l_2$, $l_3$, $c$, and $\bar c$ with 
momenta $p_1$, $p_2$, $p_3$, $p_4$, and $p_5$ to combine into a hidden-charm pentaquark state $P^{c\bar c}$,
and it should depend on the momenta of the constituents and their situated environments represented by $N_{q_i}$ and $N_{\bar{q}_i}$.
$\mathcal {R}_{l_{01}l_{02}l_{03}}$ guarantees the quark flavor conservation during the hadronization process and the product
$\mathcal {R}_{l_{01}l_{02}l_{03}} \mathcal {R}_{P^{c\bar c},l_1l_2l_3c\bar c}$ presents the probability for the $l_1$, $l_2$, $l_3$, $c$ and $\bar c$ 
to combine into a pentaquark state $P^{c\bar c}$ with the given quark contents $(l_{01}l_{02}l_{03}c\bar c)$.
So $\mathcal {R}_{l_{01}l_{02}l_{03}}$ contains the Kronecker $\delta$'s
and equals to $N_{iter}\delta_{l_1,l_{01}}\delta_{l_2,l_{02}}\delta_{l_3,l_{03}}$.
$N_{iter}$ stands for the number of possible iterations of $l_{01}l_{02}l_{03}$, and it is taken to be $1$, $3$, and $6$ for three identical 
flavor, two different flavor, and three different flavor cases, respectively. 
For example, $\mathcal {R}_{uud}=(\delta_{l_1,u}\delta_{l_2,u}\delta_{l_3,d} + 
\delta_{l_1,u}\delta_{l_2,d}\delta_{l_3,u} + \delta_{l_1,d}\delta_{l_2,u}\delta_{l_3,u})$.
$C_{P^{c\bar c}_j}$ denotes the probability for that the intrinsic quantum numbers of the formed $P^{c\bar c}$ with the quark contents $(l_{01}l_{02}l_{03}c\bar c)$
are the same as those of $P_j^{c\bar c}$.
A pentaquark state, composed of five quarks and antiquarks, every of which has spin 1/2, has three different spin quantum numbers, i.e., $J=1/2,~3/2,~5/2$.
We introduce $R_{J_{31}}$ and $R_{J_{51}}$ to denote the relative production ratio of $J=3/2$ to $J=1/2$ pentaquark states 
and that of $J=5/2$ to $J=1/2$ pentaquark states with the same flavor compositions. 
Considering that each quark has parity +1 and each antiquark has parity -1, and we only include the ground $L=0$ and the first excited $L=1$ pentaquark states,
a pentaquark state has two different parity quantum numbers, i.e., $P=-1,~+1$.
We introduce $R_{P_{10}}$ to denote the relative production ratio of $L=1$ to $L=0$ pentaquark states with the same flavor compositions. 
So we can obtain
\begin{equation}
C_{P_j^{c\bar c}} =  \left\{
\begin{array}{ll}
\frac{1}{(1+R_{J_{31}}+R_{J_{51}})} \times \frac{1}{(1+R_{P_{10}})}~~~~         \textrm{for } J^P=({1}/{2})^-  \textrm{ states} \\
\frac{1}{(1+R_{J_{31}}+R_{J_{51}})} \times \frac{R_{P_{10}}}{(1+R_{P_{10}})}~~~~         \textrm{for } J^P=({1}/{2})^+  \textrm{ states} \\
\frac{R_{J_{31}}}{(1+R_{J_{31}}+R_{J_{51}})} \times \frac{1}{(1+R_{P_{10}})}~~~~   \textrm{for } J^P=({3}/{2})^-  \textrm{ states}  \\
\frac{R_{J_{31}}}{(1+R_{J_{31}}+R_{J_{51}})} \times \frac{R_{P_{10}}}{(1+R_{P_{10}})}~~~~   \textrm{for } J^P=({3}/{2})^+  \textrm{ states}  \\
\frac{R_{J_{51}}}{(1+R_{J_{31}}+R_{J_{51}})} \times \frac{1}{(1+R_{P_{10}})}~~~~         \textrm{for } J^P=({5}/{2})^-  \textrm{ states}  \\
\frac{R_{J_{51}}}{(1+R_{J_{31}}+R_{J_{51}})} \times \frac{R_{P_{10}}}{(1+R_{P_{10}})}~~~~         \textrm{for } J^P=({5}/{2})^+  \textrm{ states}.
\end{array} \right.
\end{equation}
$R_{J_{31}}$ and $R_{J_{51}}$ have been determined to be 1.6 and 0.6 according to a simple spin counting.
$R_{P_{10}}$ has been determined to be 0.258 by the Wigner function method in Ref. \cite{ExHIC2011PRL}.
The still left unknown item is $\mathcal {R}_{P^{c\bar c},l_1l_2l_3c\bar c}(p_1,p_2,p_3,p_4,p_5;N_{q_i},N_{\bar{q}_i})$.
Recalling that it is the probability of $l_1$, $l_2$, $l_3$, $c$, and $\bar c$ with 
momenta $p_1$, $p_2$, $p_3$, $p_4$, and $p_5$ to combine into a hidden-charm pentaquark state $P^{c\bar c}$, we assume that
it is $u,~d,~s$-flavor independent, i.e.,
{\setlength\arraycolsep{0.2pt}
\begin{eqnarray}
 \mathcal {R}_{P^{c\bar c},l_1l_2l_3c\bar c}(p_1,p_2,p_3,p_4,p_5;N_{q_i},N_{\bar{q}_i})
 =\mathcal {R}_{P^{c\bar c},lllc\bar c}(p_1,p_2,p_3,p_4,p_5;N_{l},N_{\bar l},N_{q},N_{\bar q}). \label{eq:assum2}
\end{eqnarray} }%

Substitute Eqs.~(\ref{eq:Rfact}) and (\ref{eq:assum2}) into Eq.~(\ref{eq:NPccb-assum1}), we have
{\setlength\arraycolsep{0.2pt}
\begin{eqnarray}
&&  \overline{N}_{P^{c\bar c}_j}(N_{q_i},N_{\bar{q}_i})= \sum\limits_{l_{1}l_{2}l_{3}}  N_{l_1l_2l_3c\bar c}  C_{P^{c\bar c}_j}  \mathcal {R}_{l_{01}l_{02}l_{03}} 
           \nonumber     \\
&&~~~~ \times  \int dp dp_1 dp_2 dp_3 dp_4 dp_5    
f^{(n)}_{lllc\bar c}(p_1,p_2,p_3,p_4,p_5;N_l,N_{\bar l},N_q,N_{\bar q}) \mathcal{R}_{P^{c\bar c},lllc\bar c}(p_1,p_2,p_3,p_4,p_5;N_l,N_{\bar l},N_q,N_{\bar q})
     \delta(\displaystyle{\sum^5_{i=1}} p_i-p).   \label{eq:NPj2as}
\end{eqnarray} }%
We denote the momentum integral in the second line in Eq.~(\ref{eq:NPj2as}) to be $\gamma_{P^{c\bar c}}(N_{l},N_{\bar l},N_{q},N_{\bar q})$ and then obtain
{\setlength\arraycolsep{0.2pt}
\begin{eqnarray}
\overline{N}_{P^{c\bar c}_j}(N_{q_i},N_{\bar{q}_i})=  \sum\limits_{l_{1}l_{2}l_{3}}  N_{l_1l_2l_3c\bar c} 
                                                    C_{P^{c\bar c}_j}  \mathcal {R}_{l_{01}l_{02}l_{03}} 
          \gamma_{P^{c\bar c}}(N_{l},N_{\bar l},N_{q},N_{\bar q}).  \label{eq:NPj-gamma}
\end{eqnarray} }%
Assuming that the internal configurations of all different species of hidden-charm pentaquark states are the same and
summing over different species of hidden-charm pentaquark states, 
we obtain the average number of all the hidden-charm pentaquark states $\overline{N}_{P^{c\bar c}}$ as follows
{\setlength\arraycolsep{0.2pt}
\begin{eqnarray}
\overline{N}_{P^{c\bar c}}(N_{l},N_{\bar l},N_{q},N_{\bar q})  =   N_{lllc\bar c} 
          \gamma_{P^{c\bar c}}(N_{l},N_{\bar l},N_{q},N_{\bar q}),  \label{eq:NPccb-gamma}
\end{eqnarray} }%
where $N_{lllc\bar c}=N_{lll}N_{c}N_{\bar c}=N_{l}(N_{l}-1)(N_{l}-2)N_{c}N_{\bar c}$.
Substitute Eq.~(\ref{eq:NPccb-gamma}) into Eq.~(\ref{eq:NPj-gamma}), the average number of a specified hidden-charm pentaquark state $P_j^{c\bar c}$ is given by 
{\setlength\arraycolsep{0.2pt}
\begin{eqnarray}
\overline{N}_{P^{c\bar c}_j}(N_{q_i},N_{\bar{q}_i})= && \sum\limits_{l_{1}l_{2}l_{3}}   
                                                   C_{P^{c\bar c}_j}  \mathcal {R}_{l_{01}l_{02}l_{03}}
         \frac{N_{l_1l_2l_3c\bar c}}{N_{lllc\bar c}} \overline{N}_{P^{c\bar c}}(N_{l},N_{\bar l},N_{q},N_{\bar q})
         \nonumber  \\
         =&& C_{P^{c\bar c}_j} N_{iter} \frac{N_{l_{01}l_{02}l_{03}c\bar c}}{N_{lllc\bar c}}  \overline{N}_{P^{c\bar c}}(N_{l},N_{\bar l},N_{q},N_{\bar q})
         \nonumber  \\
         =&& C_{P^{c\bar c}_j} N_{iter} \frac{N_{l_{01}l_{02}l_{03}}}{N_{lll}}  \overline{N}_{P^{c\bar c}}(N_{l},N_{\bar l},N_{q},N_{\bar q}).
         \label{eq:NPj}
\end{eqnarray} }%

For a reaction at a given energy, the average numbers of quarks of different flavors $\langle N_{q_i} \rangle$ and those of antiquarks of different
flavors $\langle N_{\bar{q}_i} \rangle$ are fixed while $N_{q_i}$ and $N_{\bar{q}_i}$ follow a certain fluctuation distribution.
In this work, we focus on the midrapidity region at high LHC energy where the influence of net quarks from the colliding nuclei is negligible \cite{pikpPRL2012}.
We suppose a polynomial distribution for both the numbers of $u$, $d$ and $s$ quarks at a given $N_l$ and the numbers of $\bar u$, $\bar d$ and $\bar s$ 
at a given $N_{\bar{l}}$ with the prior probabilities $p_u=p_d=p_{\bar u}=p_{\bar d}=1/(2+\lambda_s)$, $p_s=p_{\bar s}=\lambda_s/(2+\lambda_s)$. 
Here, we introduce $\lambda_s$ to denote the production suppression of strange quarks.
Averaging over this distribution, Eq.~(\ref{eq:NPj}) becomes
{\setlength\arraycolsep{0.2pt}
\begin{eqnarray}
\overline{N}_{P^{c\bar c}_j}(N_l,N_{\bar l},N_q,N_{\bar q})
         = C_{P^{c\bar c}_j} N_{iter} p_{l_{01}}p_{l_{02}}p_{l_{03}}  \overline{N}_{P^{c\bar c}}(N_{l},N_{\bar l},N_{q},N_{\bar q}),
         \label{eq:NPj-pli}
\end{eqnarray} }%

We will also consider the fluctuations of $N_{l}$, $N_{\bar l}$, $N_{q}$ and $N_{\bar q}$ in the given kinematic region. 
By averaging over this fluctuation distribution with the
fixed $\langle N_{l} \rangle$, $\langle N_{\bar l}, \rangle$ $\langle N_{q} \rangle$ and $\langle N_{\bar q} \rangle$, we have
{\setlength\arraycolsep{0.2pt}
\begin{eqnarray}
\langle N_{P^{c\bar c}_j}\rangle(\langle N_l\rangle,\langle N_{\bar l}\rangle,\langle N_q\rangle,\langle N_{\bar q}\rangle)
         = C_{P^{c\bar c}_j} N_{iter} p_{l_{01}}p_{l_{02}}p_{l_{03}}
         \langle N_{P^{c\bar c}}\rangle(\langle N_l\rangle,\langle N_{\bar l}\rangle,\langle N_q\rangle,\langle N_{\bar q}\rangle),
         \label{eq:NPj-ave}
\end{eqnarray} }%
where $\langle N_{P^{c\bar c}}\rangle$ stands for the total average number of all the hidden-charm pentaquark states produced in the combination process. 
Eq.~(\ref{eq:NPj-ave}) is the general yield formulism of hidden-charm pentaquark states obtained from the basic ideas of the QCM with three assumptions,
i.e., the $u,~d,~s$ flavor independence of the normalized quark/antiquark joint momentum distribution,
the factorization of the kernel function and the $u,~d,~s$ flavor independence of the momentum dependent part $\mathcal{R}_{P^{c\bar c},lllc\bar c}$,
and the same configuration for all different species of hidden-charm pentaquark states.

\subsection{Yield ratios of different hidden-charm pentaquark states}

From the yield formula in Eq.~(\ref{eq:NPj-ave}), we see that there exist many simple relationships between the yields of different hidden-charm pentaquark states.
It is necessary to stress that these relations are the general features of the QCM under the three assumptions mentioned in the last subsection. 
Actrally, they are independent of the detailed form of the momentum dependence of the kernel function or the momentum distributions of the quarks and antiquarks.
They even do not depend on whether other exotic states, such as tetraquark states, pentaquark states in light sectors and so on, are produced during the 
hadronization process. They are the characteristics for hidden-charm pentaquark resonance production in the QCM, which can be used to test the 
combination production mechanism of the exotic pentaquark states.
We will propose some of these interesting relations in the following.
The first group concerns hidden-charm pentaquark states with the same spin and parity quantum numbers, and they are listed as follows
{\setlength\arraycolsep{0.2pt}
\begin{eqnarray}
 &&  \frac{\langle N_{P^{c\bar c}(uudc\bar c)}\rangle} {\langle N_{P^{c\bar c}(uddc\bar c)}\rangle} = 1,   ~~~~~~~~~
      \frac{\langle N_{P^{c\bar c}(uuuc\bar c)}\rangle} {\langle N_{P^{c\bar c}(dddc\bar c)}\rangle} = 1,    \label{eq:ratio-1}    \\
 &&  \frac{\langle N_{P^{c\bar c}(uusc\bar c)}\rangle} {\langle N_{P^{c\bar c}(ddsc\bar c)}\rangle} = 1,   ~~~~~~~~~
      \frac{\langle N_{P^{c\bar c}(ussc\bar c)}\rangle} {\langle N_{P^{c\bar c}(dssc\bar c)}\rangle} = 1,    \label{eq:ratio-2}    \\
 &&  \frac{\langle N_{P^{c\bar c}(udsc\bar c)}\rangle} {\langle N_{P^{c\bar c}(uusc\bar c)}\rangle} = 2,    ~~~~~~~~~ 
     \frac{\langle N_{P^{c\bar c}(uudc\bar c)}\rangle} {\langle N_{P^{c\bar c}(uuuc\bar c)}\rangle} = 3.     \label{eq:ratio-3}    
\end{eqnarray} }%
In the above the quark contents in the parentheses denote the flavor ingredients of the corresponding pentaquark states.
These ratios are independent of the collision energy, the species of the colliding nuclei and the collision centrality. 
Therefore, they can be used to test the universality of the production mechanism of four quarks and an antiquark combining into a pentaquark state.
In addition, they clearly exhibit relative production weights for different pentaquark states,
which can be used by the experimental search for more pentaquark states. 
For example, if the resonances observed by the LHCb Collaboration are indeed pentaquark states with $(uudc\bar c)$ quark contents 
and the hidden-charm pentaquark states are indeed produced via the quark combination,
we have no reason to say there are no $P^{c\bar c}(uddc\bar c)$ produced even it has not been observed simultaneously.

The second group is related with the strangeness production denoted by $\lambda_s$, and they are as
{\setlength\arraycolsep{0.2pt}
\begin{eqnarray}
 && \frac{\langle N_{P^{c\bar c}(uusc\bar c)}\rangle} {\langle N_{P^{c\bar c}(uudc\bar c)}\rangle} = \lambda_s,   ~~~~~~~~~
     \frac{\langle N_{P^{c\bar c}(ussc\bar c)}\rangle} {\langle N_{P^{c\bar c}(uudc\bar c)}\rangle} = \lambda_s^2,    \label{eq:ratio-lams1}  \\
 && \frac{\langle N_{P^{c\bar c}(ussc\bar c)}\rangle} {\langle N_{P^{c\bar c}(uusc\bar c)}\rangle} = \lambda_s,   ~~~~~~~~~
     \frac{\langle N_{P^{c\bar c}(sssc\bar c)}\rangle} {\langle N_{P^{c\bar c}(uusc\bar c)}\rangle} = \frac{\lambda_s^2}{3},    \label{eq:ratio-lams2}  \\
 && \frac{\langle N_{P^{c\bar c}(sssc\bar c)}\rangle} {\langle N_{P^{c\bar c}(ussc\bar c)}\rangle} = \frac{\lambda_s}{3},  ~~~~~~~~~ 
     \frac{\langle N_{P^{c\bar c}(sssc\bar c)}\rangle} {\langle N_{P^{c\bar c}(udsc\bar c)}\rangle} = \frac{\lambda_s^2}{6},       \label{eq:ratio-lams3}  \\
 &&  \frac{\langle N_{P^{c\bar c}(udsc\bar c)}\rangle} {\langle N_{P^{c\bar c}(uudc\bar c)}\rangle} = 2\lambda_s,  ~~~~~~~~
     \frac{\langle N_{P^{c\bar c}(udsc\bar c)}\rangle} {\langle N_{P^{c\bar c}(uuuc\bar c)}\rangle} = 6\lambda_s     .       \label{eq:ratio-lams} 
\end{eqnarray} }%
These ratios are also the general results of the quark combination with the above three assumptions,
and they show the relative production weights of strange hidden-charm pentaquark states to those without strangeness. 
Generally speaking, the production of strange hidden-charm pentaquark states is suppressed relative to non-strange ones due to 
the strangeness production suppression.
This just shows the advantages of heavy ion collisions to produce strange hidden-charm pentaquark states compared to elementary particle reactions
because of the strangeness enhancement in high energy heavy ion collisions.
More interestingly, from Eq.~(\ref{eq:ratio-lams}) one can see that 
strange $P^{c\bar c}(udsc\bar c)$ is nearly not suppressed relative to $P^{c\bar c}(uudc\bar c)$ 
and instead enhanced by a factor of about three relative to the non-strange $P^{c\bar c}(uuuc\bar c)$ 
considering that $\lambda_s$ in heavy ion collisions is located in the range (0.4-0.5) \cite{RQWang2012PRC}.
These results provide important references for future experimental search for strange hidden-charm pentaquark states.

As a brief summary of Sec. II, we want to emphasize once more that the method we consider in this section is intended to be a general case 
based on the basic ideas of the QCM. To simplify the results, we adopt a few assumptions and/or approximations based on symmetry and general principles,
such as the $u,~d,~s$ flavor independence of the normalized quark/antiquark joint momentum distribution,
the factorization of the kernel function and the $u,~d,~s$ flavor independence of the momentum dependent part $\mathcal{R}_{P^{c\bar c},lllc\bar c}$,
and the same configuration for all different species of hidden-charm pentaquark states.
Except the assumptions and/or approximations stated explicitly in the work, the results do not depend on other factors.
They do not depend on the specific collision energy, colliding nuclei or the centrality.
They even are not limited to heavy ion collisions, but can be suitable in elementary particle reactions if the reaction energy is large enough
to create a relative bulk quark-antiquark system, e.g., possibly in LHC $p+p$ reactions.
Therefore, these results on yield ratios can be used to test the combination production mechanism of the pentaquark states.
We focus on different hidden-charm pentaquark state yield ratios in this section. No efforts are made to study the yields, and we leave them to
the next section.

\section{Estimates for the yields of hidden-charm pentaquark states} 

The yields of various exotic resonances are the most foundmental quantities and valuable probes for exploring their internal configurations and their production 
mechanisms \cite{Lee2008EPJC,ExHIC2011PRL,ExHIC2011PRC,LWChen2007PRC,Lee2015EPJC}. 
Phenomenological models such as the statistical model and the coalescence model have been used to predict the yields for different possible exotic states,
e.g., $f_0(980)$, $a_0(980)$, $X(3872)$, $D_{sJ}(2317)$, $\varTheta_{cs}(uuds\bar c)$, .etc., \cite{Lee2008EPJC,ExHIC2011PRL,ExHIC2011PRC,LWChen2007PRC}.
In this section, we will evaluate the yields for different species of the hidden-charm pentaquark resonances produced in heavy ion collisions at the LHC.

Recalling Eq.~(\ref{eq:NPj2as}), the average number of the produced $P^{c\bar c}_j$ can be written after integrating over $p$ as follows
{\setlength\arraycolsep{0.2pt}
\begin{eqnarray}
 \overline{N}_{P^{c\bar c}_j}(N_{q_i},N_{\bar{q}_i})= && \sum\limits_{l_{1}l_{2}l_{3}}  N_{l_1l_2l_3c\bar c}  C_{P^{c\bar c}_j}  \mathcal {R}_{l_{01}l_{02}l_{03}} 
           \nonumber     \\
&& \times  \int dp_1 dp_2 dp_3 dp_4 dp_5    
f^{(n)}_{lllc\bar c}(p_1,p_2,p_3,p_4,p_5;N_l,N_{\bar l},N_q,N_{\bar q})
\mathcal{R}_{P^{c\bar c},lllc\bar c}(p_1,p_2,p_3,p_4,p_5;N_l,N_{\bar l},N_q,N_{\bar q}).   \label{eq:NPj-integ-p}
\end{eqnarray} }%
To further simplify the formalism, we ignore the transverse momentum information and only include the longitudinal part. 
We adopt the rapidity coordinate $y$ to replace the momentum coordinate and obtain
{\setlength\arraycolsep{0.2pt}
\begin{eqnarray}
 \overline{N}_{P^{c\bar c}_j}(N_{q_i},N_{\bar{q}_i})= && \sum\limits_{l_{1}l_{2}l_{3}}  N_{l_1l_2l_3c\bar c}  C_{P^{c\bar c}_j}  \mathcal {R}_{l_{01}l_{02}l_{03}} 
           \nonumber     \\
&& \times  \int dy_1 dy_2 dy_3 dy_4 dy_5    
f^{(n)}_{lllc\bar c}(y_1,y_2,y_3,y_4,y_5;N_l,N_{\bar l},N_q,N_{\bar q})
\mathcal{R}_{P^{c\bar c},lllc\bar c}(y_1,y_2,y_3,y_4,y_5;N_l,N_{\bar l},N_q,N_{\bar q}).   \label{eq:NPj-y}
\end{eqnarray} }%
Confining ourselves to the hadron production at the midrapidity $y\in (-0.5,0.5)$, we set a uniform quark joint distribution, i.e.,
$f^{(n)}_{lllc\bar c}(y_1,y_2,y_3,y_4,y_5;N_l,N_{\bar l},N_q,N_{\bar q}) = 1$. The item $\mathcal{R}_{P^{c\bar c},lllc\bar c}$
has no analytical form currently, so we rely on a specific Quark Combination Model developed by ShanDong group SDQCM \cite{QBXie1988PRD,CEShao2009PRC}.

In the following, we first give a brief introduction to the old version of the SDQCM \cite{FLShao2005PRC,QBXie1988PRD,CEShao2009PRC} and show 
how to extend the old version to include the production of the exotic multiquark states. 
Then we use the SDQCM (old version and new version) to calculate the yields of the normal identified light, strange and charm hadrons.
Finally, we estimate the yields of hidden-charm pentaquark states.
The results in central Pb+Pb collisions at $\sqrt{s_{NN}}=2.76$ TeV are given as an example.

\subsection{An introduction to the SDQCM}   \label{SDQCM}

The starting point of SDQCM is a color singlet system that consists of constituent quarks and antiquarks. 
All of these quarks and antiquarks are lined up in phase space, e.g., in a one-dimensional rapidity axis,
 and then combine into different initial hadrons one by one according to the following quark combination rule.
 \begin{enumerate}[(i)]
    \item Start from the first parton ($q_1$ or $\bar q_1$) in the line.
    \item If the baryon number of the second parton in the line is different from that of the first, i.e., the first two partons are either $q_1\bar q_2$ or $\bar q_1q_2$, 
          they combine into a meson and are removed from the line; then go back to point (i). If they are either $q_1q_2$ or $\bar q_1\bar q_2$, then go to the next point.
    \item Look at the third parton, if its baryon number is the same as the first two (i.e., $q_1q_2q_3$ or $\bar q_1\bar q_2\bar q_3$), the first three partons combine into a baryon or an antibaryon and are removed from the line; 
          then go back to point (i). Otherwise, (i.e., $q_1q_2\bar q_3$ or $\bar q_1\bar q_2 q_3$), 
          considering the color factor for quark-antiquark pairs in the color singlet channel is about twice larger than that for diquarks in the 
          color antitriplet, $q_1\bar q_3$ or $\bar q_1q_3$ form a meson and are removed from the line;
          $q_2$ or $\bar q_2$ is still left as the first parton in the line; then go back to point (i).
          
 \end{enumerate}
Here gives an example to show how the above quark combination rule works
{\setlength\arraycolsep{0.2pt}
\begin{eqnarray}
&& q_1\bar q_2\bar q_3\bar q_4\bar q_5\bar q_6\bar q_7q_8q_9\bar q_{10}q_{11}q_{12}q_{13}q_{14}q_{15}q_{16}\bar q_{17}q_{18}q_{19}\cdot\cdot\cdot\cdot\cdot\cdot \nonumber \\
 \rightarrow && M(q_1\bar q_2) \bar B(\bar q_3\bar q_4\bar q_5) M(\bar q_6q_8) M(\bar q_7q_9)  M(\bar q_{10}q_{11}) B(q_{12}q_{13}q_{14}) M(q_{15}\bar q_{17}) B(q_{16}q_{18}q_{19})\cdot\cdot\cdot\cdot\cdot\cdot.
 \label{eq:QCRold}
\end{eqnarray} }%
After applying the quark combination rule in the quark-antiquark system, the numbers of mesons, baryons and antibaryons with given quark contents
are determined.
Including the production weights of different hadrons with the same quark contents which are determined by their intrinsic quantum numbers such as spin and parity,
the multiplicities of various kinds of initial hadrons are obtained. 
A more detailed description for the old version of the SDQCM can be found in Refs. \cite{FLShao2005PRC,QBXie1988PRD,CEShao2009PRC}.

The above old version of SDQCM sets the priority of the smallest number of partons to form a hadron, and there are no multiquark states produced. 
In the following, we will extend the quark combination rule to include exotic multiquark states. 
Considering the large difference between heavy charm quarks and $u,~d,~s$ quarks, the previous quark combination rule is extended to be a new one as follows.

 \begin{enumerate}[(i)]
    \item Start from the first parton ($q_1$ or $\bar q_1$) in the line.
    \item If the baryon number of the second parton in the line is different from that of the first, i.e., the first two partons are either $q_1\bar q_2$ or $\bar q_1q_2$, 
          they combine into a meson and are removed from the line; then go back to point (i). If they are either $q_1q_2$ or $\bar q_1\bar q_2$, then go to point (iii).
    \item Look at the third parton, if its baryon number is the same as the first two, (i.e., $q_1q_2q_3$ or $\bar q_1\bar q_2\bar q_3$), the first three partons combine into a baryon or an antibaryon and are removed from the line; 
          then go back to point (i). 
          Otherwise, (i.e., $q_1q_2\bar q_3$ or $\bar q_1\bar q_2 q_3$), if the first and the third partons are both heavy charm flavors or light/strange flavors,
          $q_1\bar q_3$ or $\bar q_1q_3$ form a meson and are removed from the line;
          $q_2$ or $\bar q_2$ is still left in the line; then go back to point (i).
          If one of the first and the third partons is in charm flavor and the other is in light/strange flavors, 
          the probability for them to combine into a meson will decrease due to 
          their relatively large momentum difference resulted from their large mass difference \cite{Greco2003}.
          If they can not combine into $D$ mesons, multiquark states are the most possible candidates for them to contribute to.
          We use $\varepsilon$ to denote the probability for the first and the third partons to combine with other nearby partons to
          merge into multiquark states, and in this case, go to point (iv).         
    \item Look at the fourth parton, if its baryon number is the same as the third, (i.e., $q_1q_2\bar q_3\bar q_4$ or $\bar q_1\bar q_2 q_3q_4$),
          the first four partons combine into a tetraquark state, and then go back to point (i). Otherwise, (i.e., $q_1q_2\bar q_3q_4$ or $\bar q_1\bar q_2 q_3\bar q_4$),
          go to point (v).
    \item Look at the fifth parton, if its baryon number is the same as the fourth, (i.e., $q_1q_2\bar q_3q_4q_5$ or $\bar q_1\bar q_2 q_3\bar q_4\bar q_5$),
          the first five partons combine into a pentaquark state, and then go back to point (i). Otherwise, (i.e., $q_1q_2\bar q_3q_4\bar q_5$ or $\bar q_1\bar q_2 q_3\bar q_4q_5$),
          go to point (vi). 
    \item Look at the sixth parton, if its baryon number is the same as the fifth, (i.e., $q_1q_2\bar q_3q_4\bar q_5\bar q_6$ or $\bar q_1\bar q_2 q_3\bar q_4q_5q_6$), 
          the first six partons combine into a sixquark state, and then go back to point (i). 
          Otherwise, continue to look at the next parton or the next next parton until they can be in a color-singlet multiquark state.
 \end{enumerate}
In the following, we give an example to show how the above new quark combination rule works
{\setlength\arraycolsep{0.2pt}
\begin{eqnarray}
&& q_1\bar q_2\bar q_3\bar q_4\bar q_5\bar l_6\bar l_7l_8l_9\bar q_{10}q_{11}q_{12}q_{13}q_{14}l_{15}q_{16}\bar cq_{18}q_{19}\cdot\cdot\cdot\cdot\cdot\cdot \nonumber \\
 \rightarrow && \left\{
\begin{array}{ll} 
 M(q_1\bar q_2) \bar B(\bar q_3\bar q_4\bar q_5) M(\bar l_6l_8) M(\bar l_7l_9)  M(\bar q_{10}q_{11}) B(q_{12}q_{13}q_{14}) M(l_{15}\bar c) B(q_{16}q_{18}q_{19})\cdot\cdot\cdot\cdot\cdot\cdot~~~ \textrm{with probability }1-\varepsilon \\
 M(q_1\bar q_2) \bar B(\bar q_3\bar q_4\bar q_5) M(\bar l_6l_8) M(\bar l_7l_9)  M(\bar q_{10}q_{11}) B(q_{12}q_{13}q_{14}) P(l_{15}q_{16}\bar cq_{18}q_{19})\cdot\cdot\cdot\cdot\cdot\cdot~~~ \textrm{with probability }\varepsilon
 .
 \label{eq:QCRnew}
 \end{array} \right.
\end{eqnarray} }%
Note that when $\varepsilon=0$, the new quark combination rule becomes to be the old one, and when $\varepsilon=1$, the multiquark states
are allowed to be produced most abundantly.
Applying the new quark combination rule in the quark-antiquark system, the numbers of mesons, baryons, antibaryons and multiquark states with given quark contents
are determined. Including the production weights of different hadrons, the multiplicities of various kinds of initial hadrons are obtained. 

We want to state that the new quark combination rule includes the production of the multiquark states such as the tetraquark states, the 
pentaquark states, the sixquark states and so on in the charm sector. 
The production of the multiquark states in light and strange sectors are simply neglected because the $q_1q_2\bar q_3$ 
or $\bar q_1\bar q_2q_3$ with light and/or strange flavors in the line are easily
to form mesons. In addition, the production of the multiquark states with more valence quarks and/or antiquarks is
suppressed more strongly in the new quark combination rule, which follows the general principles during the hadron production.
The quark combination rule in the SDQCM naturally satisfy the near phase space correlation for quarks and/or antiquarks when they combine into different hadrons
and the unitary requirement simultaneously.
Also, it determines the competition of the hadron production uniquely.
Although very simple for the quark combination rule, it can meet the constraints for the kernel functions of hadrons
and make the calculations possible.

\subsection{Yields of normal identified hadrons}

\begin{table*}[htbp]
\renewcommand{\arraystretch}{1.5}
\caption{Midrapidity yields $dN/dy$ of normal identified mesons and baryons in central Pb+Pb collisions at $\sqrt{s_{NN}}=2.76$ TeV.
The experimental data are from Refs. \cite{pikpPRL2012,Ks0lam2013,xiome2013,phipt}.}
\begin{tabular}{p{60pt}p{90pt}p{80pt}p{80pt}p{70pt}}
\toprule
 Hadron                                      & Data                &SDQCM ($\varepsilon=0$) &SDQCM ($\varepsilon=1$) &SDQCM ($\varepsilon=0.6$)    \\
\colrule
$\pi^+$                                      &$733\pm54$           &$742$                   &$728$                   &$734$      \\
$\pi^-$                                      &$732\pm52$           &$742$                   &$728$                   &$734$      \\
$K^+$                                        &$109\pm9$            &$115$                   &$111$                   &$112$     \\
$K^-$                                        &$109\pm9$            &$115$                   &$111$                   &$112$     \\
$K_S^{0}$                                    &$110\pm10$           &$109$                   &$106$                   &$107$     \\
$\phi$                                       &$13.8\pm0.5\pm1.7$   &$14.8$                  &$14.5$                  &$14.6$   \\
$p$                                          &$34\pm3$             &$32$                    &$31$                    &$32$    \\
$\bar p$                                     &$32\pm3$             &$32$                    &$31$                    &$32$    \\  
$\Lambda$                                    &$26\pm3$             &$25$                    &$25$                    &$25$    \\
$\bar\Lambda$                                &---                  &$25$                    &$25$                    &$25$    \\ 
$\Xi^-$                                      &$3.34\pm0.06\pm0.24$ &$3.89$                  &$3.84$                  &$3.86$    \\
$\bar\Xi^+$                                  &$3.28\pm0.06\pm0.23$ &$3.89$                  &$3.84$                  &$3.86$    \\  
$\Omega^-$                                   &$0.58\pm0.04\pm0.09$ &$0.55$                  &$0.54$                  &$0.54$     \\
$\bar\Omega^+$                               &$0.60\pm0.05\pm0.09$ &$0.55$                  &$0.54$                  &$0.54$     \\
\colrule
$D^+$ ($D^-$)                                &---                  &$3.35$                  &$2.29$                  &$2.71$    \\
$D^0$ ($\bar D^0$)                           &---                  &$10.4$                  &$7.10$                  &$8.41$  \\
$D^{*+}$ ($D^{*-}$)                          &---                  &$5.15$                  &$3.52$                  &$4.17$   \\
$D^{*0}$ ($\bar D^{*0}$)                     &---                  &$5.15$                  &$3.52$                  &$4.17$   \\
$D_s^+$ ($D_s^-$)                            &---                  &$2.82$                  &$1.93$                  &$2.28$   \\
$D_s^{*+}$ ($D_s^{*-}$)                      &---                  &$2.11$                  &$1.44$                  &$1.71$     \\
$\eta_c$                                     &---                  &$0.054$                 &$0.055$                 &$0.054$   \\
$J/\Psi$                                     &---                  &$0.160$                 &$0.166$                 &$0.164$     \\
$\Lambda_c^+$ ($\bar\Lambda_c^-$)            &---                  &$2.86$                  &$2.83$                  &$2.84$    \\
$\Sigma_c^+$ ($\bar\Sigma_c^-$)              &---                  &$0.569$                 &$0.564$                 &$0.566$    \\
$\Sigma_c^0$ ($\bar\Sigma_c^0$)              &---                  &$0.473$                 &$0.469$                 &$0.471$    \\
$\Sigma_c^{++}$ ($\bar\Sigma_c^{--}$)        &---                  &$0.474$                 &$0.470$                 &$0.471$     \\
$\Xi_{cc}^{++}$ ($\bar\Xi_{cc}^{--}$)        &---                  &$0.0140$                &$0.0139$                &$0.0140$  \\
$\Xi_{cc}^{+}$ ($\bar\Xi_{cc}^{-}$)          &---                  &$0.0140$                &$0.0139$                &$0.0140$     \\
$\Omega_{cc}^{+}$ ($\bar\Omega_{cc}^{-}$)    &---                  &$0.00580$               &$0.00577$               &$0.00574$   \\
$\Omega_{ccc}^{++}$ ($\bar\Omega_{ccc}^{--}$)&---                  &$0.000195$              &$0.000192$              &$0.000193$   \\
\botrule
\end{tabular} \label{tab:Nhadron}
\end{table*}

In this subsection, we use the SDQCM to compute the midrapidity yields $dN/dy$ of normal identified mesons and baryons in 
light, strange and charm sectors in central Pb+Pb collisions at $\sqrt{s_{NN}}=2.76$ TeV.
Before doing that, we have to determine the input parameters of the SDQCM. 
The first is the number of all the midrapidity quarks and antiquarks $dN_{q+\bar q}/dy$ which determines the total multiplicity at midrapidity 
and is related with the reaction energy, the species of the collision nuclei, the collision centrality, the studied dynamical space, and so on.
In this paper, we set $dN_{q+\bar q}/dy=2dN_q/dy=2dN_{\bar q}/dy=3300$, 
where the net quark number is set to be zero due to focusing on the midrapidity region at so high LHC energy \cite{pikpPRL2012}.
The second input parameter is the strangeness suppression factor $\lambda_s$ that determines the number of the strange quarks.
We adopt the saturated value for $\lambda_s$ in relativistic heavy ion collisions, i.e., $\lambda_s=0.41$ \cite{RQWang2015PRC}. 
The last input parameter is the number of charm quarks in the quark-antiquark system.
We set it by extrapolating $p+p$ reaction data at LHC as follows
 \begin{equation}
   \frac{dN_{c}}{dy}=\langle T_{AA}\rangle \frac{d\sigma^{pp}_{c}}{dy}=\langle T_{AA}\rangle \frac{1}{R}\frac{d\sigma^{pp}_{D^0}}{dy}=21. 
 \end{equation}
Here $R=0.54\pm0.05$ is the branch ratio of charm quarks into final $D^0$ mesons measured in $e^+e^-$ reactions \cite{D02005PRL}.
$\langle T_{AA}\rangle=26.4\pm0.5~\mathrm{mb}^{-1}$ is the average nuclear overlap function in the most central (0-5\% centrality) 
Pb+Pb collisions at $\sqrt{s_{NN}}=2.76$ TeV calculated with the Glauber model \cite{HpmPLB2011}.
The differential cross section of $D^0$ is ${d\sigma_{D^0}^{pp}}/{dy}=0.428\pm0.115$ mb in $p+p$ reactions at $\sqrt{s}= 2.76$ TeV \cite{D0crosssec}.

The calculated midrapidity yields $dN/dy$ of different identified hadrons are collected in Table~\ref{tab:Nhadron}. 
The contributions from strong and electromagnetic decays for light and strange hadrons have been included to coincide with the ALICE experiments.
For charmed hadrons, we only consider the decays of $D^*$ mesons for $D$ mesons and $\Sigma_c$ and $\Sigma_c^*$ baryons for $\Lambda_c^+$.
The third and the fourth columns in Table~\ref{tab:Nhadron} are the results calculated by the old version of the SDQCM in which there are no 
multiquark states included and those by the new version of the SDQCM with $\varepsilon=1$ where the multiquark states are allowed to be 
produced most abundantly.
From Table~\ref{tab:Nhadron}, one can see that the results for light and strange hadrons in these two limit cases, i.e., $\varepsilon=0$ and $\varepsilon=1$,
are nearly the same and they both agree well with the available data from Refs. \cite{pikpPRL2012,Ks0lam2013,xiome2013,phipt}.
For charmed baryons and hidden-charm mesons, the results with $\varepsilon=1$ and $\varepsilon=0$ are comparable.
For open charmed mesons, the results with $\varepsilon=1$ decrease about 30\% compared to those with $\varepsilon=0$.
The predicted yield of $\Omega^{++}_{ccc}$ is in good agreement with the result from the coalescence model \cite{PFZhuang2015PLB}.
The computed results for the other charmed hadrons wait for the comparisons with the future experimental measurements
and/or other theoretical calculations.

\subsection{Estimates for the yields of various hidden-charm pentaquark states}

Based on the success in the description of the yields of normal light and strange  hadrons, we will in this subsection employ SDQCM to 
give an estimate for the midrapidity yields of different hidden-charm pentaquark states in the most central Pb+Pb collisions at $\sqrt{s_{NN}}=2.76$ TeV.
Setting $\varepsilon=1$, we obtain the up limit value for the number of all the hidden-charm pentaquark 
states $dN_{P^{c\bar c}}/dy=0.0411$.
Taking it into Eq.~(\ref{eq:NPj-ave}), we can predict the up limit values for the multiplicities of different hidden-charm pentaquark states.
The results are in Table \ref{tab:NpentaUPL}.

\begin{table*}[htbp]
\renewcommand{\arraystretch}{2.5}
\caption{Up limit values for the midrapidity yields $dN/dy$ of various hidden-charm pentaquark states with different spins and parities in the most central Pb+Pb collisions at $\sqrt{s_{NN}}=2.76$ TeV.}
\begin{tabular}{@{}c|cccccc|cccccc@{}}
\toprule
Flavor  &\multicolumn{6}{c|}{$uudc\bar c$}                                      &\multicolumn{6}{c}{$uddc\bar c$}           \\
$J^P$   &$(1/2)^-$  &$(1/2)^+$   &$(3/2)^-$ &$(3/2)^+$   &$(5/2)^-$ &$(5/2)^+$     &$(1/2)^-$  &$(1/2)^+$   &$(3/2)^-$ &$(3/2)^+$   &$(5/2)^-$ &$(5/2)^+$                 \\
$\frac{dN}{dy}~~(10^{-4})$ &21.9  &5.65  &35.0  &9.03  &13.1  &3.39            &21.9  &5.65  &35.0  &9.03  &13.1  &3.39  \\
\colrule
Flavor  &\multicolumn{6}{c|}{$uuuc\bar c$}                                      &\multicolumn{6}{c}{$dddc\bar c$}           \\
$J^P$   &$(1/2)^-$  &$(1/2)^+$   &$(3/2)^-$ &$(3/2)^+$   &$(5/2)^-$ &$(5/2)^+$     &$(1/2)^-$  &$(1/2)^+$   &$(3/2)^-$ &$(3/2)^+$   &$(5/2)^-$ &$(5/2)^+$                 \\
$~~\frac{dN}{dy}~~(10^{-4})~~$ &7.29  &1.88  &11.7  &3.01  &4.38  &1.13            &7.29  &1.88  &11.7  &3.01  &4.38  &1.13 \\
\colrule
Flavor  &\multicolumn{6}{c|}{$uusc\bar c$}                                      &\multicolumn{6}{c}{$ddsc\bar c$}           \\
$J^P$   &$(1/2)^-$  &$(1/2)^+$   &$(3/2)^-$ &$(3/2)^+$   &$(5/2)^-$ &$(5/2)^+$     &$(1/2)^-$  &$(1/2)^+$   &$(3/2)^-$ &$(3/2)^+$   &$(5/2)^-$ &$(5/2)^+$                 \\
$~~\frac{dN}{dy}~~(10^{-4})~~$ &8.97  &2.31  &14.4  &3.70  &5.38  &1.39          &8.97  &2.31  &14.4  &3.70  &5.38  &1.39 \\
\colrule
Flavor  &\multicolumn{6}{c|}{$ussc\bar c$}                                      &\multicolumn{6}{c}{$dssc\bar c$}           \\
$J^P$   &$(1/2)^-$  &$(1/2)^+$   &$(3/2)^-$ &$(3/2)^+$   &$(5/2)^-$ &$(5/2)^+$     &$(1/2)^-$  &$(1/2)^+$   &$(3/2)^-$ &$(3/2)^+$   &$(5/2)^-$ &$(5/2)^+$                 \\
$~~\frac{dN}{dy}~~(10^{-4})~~$ &3.68  &0.949  &5.89  &1.52  &2.21  &0.569          &3.68  &0.949  &5.89  &1.52  &2.21  &0.569 \\
\colrule
Flavor  &\multicolumn{6}{c|}{$udsc\bar c$}                                      &\multicolumn{6}{c}{$sssc\bar c$}           \\
$J^P$   &$(1/2)^-$  &$(1/2)^+$   &$(3/2)^-$ &$(3/2)^+$   &$(5/2)^-$ &$(5/2)^+$     &$(1/2)^-$  &$(1/2)^+$   &$(3/2)^-$ &$(3/2)^+$   &$(5/2)^-$ &$(5/2)^+$                 \\
$~~\frac{dN}{dy}~~(10^{-4})~~$ &17.9  &4.63  &28.7  &7.41  &10.8  &2.78            &0.503  &0.130  &0.804  &0.208  &0.302  &0.0778 \\
\botrule
\end{tabular} \label{tab:NpentaUPL}
\end{table*}

To further evaluate the yields of different hidden-charm pentaquark states, we have to determine $\varepsilon$.
Here we use the experimental data on
the $p_T$ distribution of $D^0$ mesons in the 0-10\% centrality measured by the ALICE Collaboration \cite{D0pt2015}. 
We use the Blast Wave Model \cite{BWmodel} to extract the midrapidity yield for $D^0$ mesons as about $dN/dy=7.42$.
Simply scaled by the $\langle T_{AA}\rangle=26.4\pm0.5~\mathrm{mb}^{-1}$ \cite{HpmPLB2011} and $\langle T_{AA}\rangle=23.44\pm0.76~\mathrm{mb}^{-1}$ \cite{D0pt2015} in the 0-5\% centrality and 0-10\% centrality, respectively,
we have the experimental result $dN/dy\approx 8.36$ for $D^0$ mesons in the most central 0-5\% collisions.
By reproducing $D^0$ mesons, $\varepsilon$ is set to be 0.6.
The midrapidity yields for other normal identified hadrons when $\varepsilon=0.6$ are also listed in Table~\ref{tab:Nhadron}.
With $\varepsilon=0.6$, we obtain
the estimated number of all the hidden-charm pentaquark states is $dN_{P^{c\bar c}}/dy=0.0248$.
Taking it into Eq.~(\ref{eq:NPj-ave}), we can predict the multiplicities for different hidden-charm pentaquark states.
The results are in Table \ref{tab:NpentaSP}.

\begin{table*}[htbp]
\renewcommand{\arraystretch}{2.5}
\caption{Estimates for the midrapidity yields $dN/dy$ of various pentaquark states with different spins and parities in the most central Pb+Pb collisions at $\sqrt{s_{NN}}=2.76$ TeV.}
\begin{tabular}{@{}c|cccccc|cccccc@{}}
\toprule
Flavor  &\multicolumn{6}{c|}{$uudc\bar c$}                                      &\multicolumn{6}{c}{$uddc\bar c$}           \\
$J^P$   &$(1/2)^-$  &$(1/2)^+$   &$(3/2)^-$ &$(3/2)^+$   &$(5/2)^-$ &$(5/2)^+$     &$(1/2)^-$  &$(1/2)^+$   &$(3/2)^-$ &$(3/2)^+$   &$(5/2)^-$ &$(5/2)^+$                 \\
$\frac{dN}{dy}~~(10^{-4})$ &13.2  &3.41  &21.1  &5.45  &7.92  &2.04            &13.2  &3.41  &21.1  &5.45  &7.92  &2.04  \\
\colrule
Flavor  &\multicolumn{6}{c|}{$uuuc\bar c$}                                      &\multicolumn{6}{c}{$dddc\bar c$}           \\
$J^P$   &$(1/2)^-$  &$(1/2)^+$   &$(3/2)^-$ &$(3/2)^+$   &$(5/2)^-$ &$(5/2)^+$     &$(1/2)^-$  &$(1/2)^+$   &$(3/2)^-$ &$(3/2)^+$   &$(5/2)^-$ &$(5/2)^+$                 \\
$~~\frac{dN}{dy}~~(10^{-4})~~$ &4.40  &1.14  &7.04  &1.82  &2.64  &0.681         &4.40  &1.14  &7.04  &1.82  &2.64  &0.681     \\
\colrule
Flavor  &\multicolumn{6}{c|}{$uusc\bar c$}                                      &\multicolumn{6}{c}{$ddsc\bar c$}           \\
$J^P$   &$(1/2)^-$  &$(1/2)^+$   &$(3/2)^-$ &$(3/2)^+$   &$(5/2)^-$ &$(5/2)^+$     &$(1/2)^-$  &$(1/2)^+$   &$(3/2)^-$ &$(3/2)^+$   &$(5/2)^-$ &$(5/2)^+$                 \\
$~~\frac{dN}{dy}~~(10^{-4})~~$ &5.41  &1.40  &8.66  &2.23  &3.25  &0.838         &5.41  &1.40  &8.66  &2.23  &3.25  &0.838   \\
\colrule
Flavor  &\multicolumn{6}{c|}{$ussc\bar c$}                                      &\multicolumn{6}{c}{$dssc\bar c$}           \\
$J^P$   &$(1/2)^-$  &$(1/2)^+$   &$(3/2)^-$ &$(3/2)^+$   &$(5/2)^-$ &$(5/2)^+$     &$(1/2)^-$  &$(1/2)^+$   &$(3/2)^-$ &$(3/2)^+$   &$(5/2)^-$ &$(5/2)^+$                 \\
$~~\frac{dN}{dy}~~(10^{-4})~~$ &2.22  &0.573  &3.55  &0.916  &1.33  &0.344       &2.22  &0.573  &3.55  &0.916  &1.33  &0.344     \\
\colrule
Flavor  &\multicolumn{6}{c|}{$udsc\bar c$}                                      &\multicolumn{6}{c}{$sssc\bar c$}           \\
$J^P$   &$(1/2)^-$  &$(1/2)^+$   &$(3/2)^-$ &$(3/2)^+$   &$(5/2)^-$ &$(5/2)^+$     &$(1/2)^-$  &$(1/2)^+$   &$(3/2)^-$ &$(3/2)^+$   &$(5/2)^-$ &$(5/2)^+$                 \\
$~~\frac{dN}{dy}~~(10^{-4})~~$ &10.8  &2.79  &17.3  &4.47  &6.50  &1.68            &0.303  &0.0783  &0.485  &0.125  &0.182  &0.0470 \\
\botrule
\end{tabular} \label{tab:NpentaSP}
\end{table*}

Note that we neglect the contribution from the hadron-hadron rescattering in hadronic phase and assume the dominant source is from the quark combination.
The pentaquark states made by the quark contents $(uudc\bar c)$ with $J=3/2$ and $J=5/2$ in Table \ref{tab:NpentaUPL} and Table \ref{tab:NpentaSP} correspond to the observed $P_c^+(4380)$ and $P_c^+(4450)$, respectively,
from which one can see that the state with negative parity has a larger yield than that with positive parity by a factor of about four, 
indicating that the yield measurements can help to determine the parities of such observed resonances.
Our results show that the estimated yields of these different hidden-charm pentaquark states are large enough for carrying out realistic measurements.

\section{summary}

With the basic ideas of the QCM and a few assumptions and/or simplifications based on symmetry and general principles, 
we have derived the yield formulas and studied the yield ratios of different species of hidden-charm pentaquark states 
in heavy ion collisions at unltra-relativistic collision energies.
We found some interesting relations between different hidden-charm pentaquark states.
These results are properties of the pentaquark state production in the QCM under these assumptions such as 
the $u,~d,~s$ flavor independence of the normalized joint momentum distributions,
the factorization of the kernel function and the $u,~d,~s$ flavor independence of the momentum dependent part of the kernel function,
and the same internal configuration for all different species of hidden-charm pentaquark states.
They are independent of the particular quark combination models, so they can be used to test the quark
combination production mechanism of the exotic resonances.
We have also calculated the yields of normal identified light, strange and charm hadrons with the SDQCM, 
and on this basis we have shown how to estimate the values for the yields of different hidden-charm pentaquark states.
The estimated results in the most central Pb+Pb collisions at $\sqrt{s_{NN}}=2.76$ TeV are presented as an example.
All of this can help to search for the partners of the observed $P_c^+(4380)$ and $P_c^+(4450)$ 
and shed light on the understanding of the production mechanism of the exotic pentaquark states in heavy ion collisions.

\section*{Acknowledgements}

The authors thank Wei Wang, Fan Wang, Xue-Qian Li, Qiang Zhao and Xian-Hui Zhong for helpful discussions.
RQW would like to thank the members of the INPAC for their hospitalities during her stay in SJTU as a visitor.
This work is supported in part by 
the Major State Basic Research Development Program (973 Program) in China under Contract Nos. 2015CB856904 and 2013CB834405, 
the National Natural Science Foundation of China under Grant Nos. 11575100, 11575110, 11505104, 11305076, 11275125 and 11135011,
the Natural Science Foundation of Shandong Province under Grant Nos. ZR2015PA002 and ZR2015JL001,
and the Natural Science Foundation of Shanghai under Grant Nos. 15DZ2272100 and 15ZR1423100.

\end{document}